\documentclass[11pt,twoside]{article}

\usepackage{asp2004}
\usepackage{epsf}
\usepackage{psfig}
\usepackage{lscape}

\begin{document}
\title{Lightcurve Survey of V-type Asteroids. I. 
\ \newline
Observations until Spring 2004}   %%% Fill in title
\author{Sunao Hasegawa$^{1}$, Seidai Miyasaka$^{2}$, Chiaki Yoshizumi$^{3}$, 
\ \newline
Tomohiko Sekiguchi$^{4}$, Yuki Sarugaku$^{1,5}$, Setsuko Nishihara$^{1,5}$, 
\ \newline
Kouhei Kitazato$^{1,5}$, Masanao Abe$^{1}$, and Hiroyuki Mito$^{6}$}   
%%% Fill in author names
\affil{$^{1}$Institute of Space and Astronautical Science, Japan Exploration Agency, 
Kanagawa, 229-8510 JAPAN
\ \newline
$^{2}$Tokyo Metropolitan Government, Tokyo, 163-8001 JAPAN
\ \newline
$^{3}$Tokushima Science Museum, Tokushima, 779-0111 JAPAN
\ \newline
$^{4}$National Astronomical Observatory of Japan, National Institutes of Natural 
Sciences, Tokyo, 181-8588 JAPAN
\ \newline
$^{5}$Graduate School of Science, The University of Tokyo, Tokyo, 113-0033 JAPAN, 
\ \newline
$^{6}$Kiso Observatory, Institute of Astronomy, The University of Tokyo, Nagano, 
397-0101 JAPAN}    %%% Fill in author affiliations
\begin{abstract} %%% Abstract to run on from here.
To examine the distribution of rotational rates for chips of asteroid 4 Vesta, 
lightcurve observation of seven V-type asteroids (2511 Patterson, 2640 Hallstorm, 
2653 Principia, 2795 Lapage, 3307 Athabasca, 4147 Lennon, and 4977 Rauthgundis) 
were performed from fall 2003 to spring 2004.  
Distribution of spin rates of V-type main-belt asteroids from the past and our 
observations have three peaks.  
This result implies that age of catastrophic impact making Vesta family may be
not as young as Karin and Iannini families but as old as Eos and Koronis families.
\end{abstract}

%%% MAIN BODY OF TEXT GOES HERE. CONSULT "INSTRUCTIONS FOR AUTHORS USING
%%% LATEX2E MARKUP", SECTIONS 2.3-2.6 FOR HELP WITH EQUATIONS, FIGURES,
%%% AND TABLES.
%\section{}   %%% Top level section head (remove "%" symbol)
%\subsection{}   %%% Second level section head (remove "%" symbol)
%\subsubsection{}   %%% Lowest level section head (remove "%" symbol)
%\section*{}	%%% Unnumbered top level section head (remove "%" symbol)
%\subsection*{}   %%% Unnumbered second level section head (remove "%" symbol)

\section{Introduction}   %%% Top level section head (remove "%" symbol)
  It is considered that differentiated meteorites howardites, eucrites, and diogenites 
(HED meteorites) have been formed in the same regions due to the fact that the HED 
meteorites consist of basalt, cumulative gabbros, and orthopyroxenites, and have 
the same isotopic compositions.  
  McCord et al. (1970) obtained the first modern extended-visible spectrum of asteroid 
4 Vesta, and showed that the surface of Vesta has a composition similar to that of 
certain basaltic achondrites.
  It was noted that the visible and near infrared reflectance spectrum of Vesta is 
unique among the main-belt asteroids larger than 50 km and is closely matched with 
those of the HED meteorites (e.g., Larson \& Fink 1975, Feierberg \& Drake. 1980).  
  The shape of Vesta was obtained by direct imaging and a large impact crater on the 
southern hemisphere was identified (Thomas et al. 1997).  
  A number of small asteroids with Vesta-like visible spectra which are usually called 
`V-type asteroids' and/or `Vestoids' were found near Vesta orbit which was positioned 
between Vesta and the 3:1 ($\sim$2.5 AU) mean motion resonance and $\nu$6 secure 
resonance with Jupiter (Binzel and Xu 1993), and near-Earth orbit (Cruikshank, et al. 
1991).  
  Binzel et al. (2004) showed an absence of V-type asteroids among Mars crossing 
asteroids.  
  The total volume of V-type asteroids including main-belt asteroids and near-Earth 
asteroids is less than that of estimated excavation ejecta from huge crated on 
southern polar region of asteroid 4 Vesta (Thomas et al. 1997).  
  These researches have suggested that most V-type asteroids are ejected fragments 
of Vesta and one of the sources of HED meteorites is Vesta. 

%   The fact indicates the impact which made even the huge crater on Vesta is not 
%catastrophic disruption but cratering.  
%   V-type asteroids are thought to be the fragments made with the biggest cratering 
%experiment that we can observe.
%  Therefore, it is possible to bring us important clues for impact cratering in 
%gravity regime to search rotation rate of V-type asteroids. 

   From distributions of rotational rates of asteroids, we can obtain information of 
collisional age and evolution of asteroids (e.g., Dobrovolskis and Burns 1984, 
Cellino et al. 1990, Paolicchi et al. 2002, Richardson et al. 2002, Vorkrouhlicky 
et al. 2003).
   To know the collisional history of Vesta, we pay our attention to distribution 
of rotation period of V-type asteroids.
   About 75 V-type asteroids have been identified by spectroscopic method (Tholen 1984, 
Xu et al. 1995, Bus and Binzel 2002, Lazzaro et al. 2004), but the rotation properties 
of V-type asteroids are not studied enough. 
   Therefore, we have performed observations of seven new V-type asteroids.  

\section{Observations and data reduction}   %%% Top level section head (remove "%" symbol)
   The lightcurve observations of asteroids reported here were performed using three 
different telescopes in Japan.  
   The data of asteroids were obtained with the 1.05-m and the 0.30-m telescopes of the 
Kiso observatory at Nagano, Japan (MPC code 381), and the 0.25-m telescope of the 
Miyasaka Observatory at Yamanashi, Japan (MPC code 366).
   An SITe TK2048E CCD detector (2048 $\times$ 2048 pixels) giving an image scale of 
1.5$\arcsec$/pixel located in the Schmidt focus of the 1.05-m F/3.1 Schmidt telescope 
was used.
   The field of view of the CCD was 51$\arcmin$ $\times$ 51$\arcmin$.
   In the observations with the 0.30-m F/9.1 Dall-Kirkham type telescope (K.3T), 
MUTOH CV16II (Kodak KAF-1600) CCD detector having a format of 1536 $\times$ 1024 pixels 
was used with a image scale of 1.35$\arcsec$/(2 pixels) (using 2 $\times$ 2 binning), 
giving a 17.3$\arcmin$ $\times$ 11.5$\arcmin$ as sky field.
    The 0.25-m telescope was equipped with an SBIG ST-6 CCD detector whose format is 
375 $\times$ 242 pixels mounted in Newtonian forcus (F/6 system).  
   The image pixel of ST-6 CCD is 3.4$\arcsec$/pixel, yielding the field of view of 
19.7$\arcmin$ $\times$ 14.9$\arcmin$.

   Due to the fact that CV16II and ST-6 CCDs are air cooled (all observations with the 
0.25-m and a part of observations with the K.3T) or by water cooled (a part of 
observations with the K.3T), respectively. 
   Dark images are obtained in all observations of the 0.25-m telescope and the K.3T
and integration time of dark images are as the same as that used for object images.  
   Flat field images with the Schmidt telescope, the 0.25-m telescope and the K.3T 
were obtained using a white screen illuminated with one and two incandescent lamps 
and the twilight sky, respectively.  
   All observations were made through the R band filter. 
%   but the asteroid 63 Ausonia, 238 Hypatia, and 355 Gabriella were taken in the V 
%band filter for testing the lightcurve observations of two telescopes.
   The aspect data for observed asteroids are presented in Table I.

\begin{table}[!ht]
\caption{Observational Circumstances}
\smallskip
\begin{center}
{\small
\begin{tabular}{lllllll}
\tableline
\noalign{\smallskip}
Num Name & Obs date (UT) & Rh   &$\Delta$& $\alpha$ & Ba & Obs \\
         &               & [AU] &  [AU]  & [$\deg$] & nd & Tel \\
\noalign{\smallskip}
\tableline
\noalign{\smallskip}
2511 Patterson	    & 2004 Mar. 12 17:24-20:14 & 2.298 & 1.334 &  8.0 & R & Schmidt\\
(Vesta family)	    & 2004 Mar. 13 18:27-20:09 & 2.297 & 1.330 &  7.6 & R & Schmidt\\
                   & 2004 Mar. 14 11:51-17:20 & 2.296 & 1.328 &  7.4 & R & Schmidt\\
                   & 2004 Mar. 15 11:39-18:10 & 2.295 & 1.324 &  7.1 & R & Schmidt\\
                   & 2004 Mar. 15 11:34-18:31 & 2.295 & 1.324 &  7.1 & R & K.3T\\
                   & 2004 Apr. 29 11:02-18:19 & 2.242 & 1.420 & 18.9 & R & Schmidt\\
2640 Hallstrom	    & 2004 Jan. 15 17:05-20:59 & 2.067 & 1.391 & 24.3 & R & Schmidt\\
(non Vesta family) & 2004 Mar. 12 09:45-10:48 & 2.226 & 1.285 & 10.9 & R & Schmidt\\
                   & 2004 Mar. 13 09:44-18:22 & 2.224 & 1.290 & 11.5 & R & Schmidt\\
                   & 2004 Mar. 14 09:46-17:02 & 2.224 & 1.294 & 11.9 & R & Schmidt\\
                   & 2004 Mar. 15 09:47-17:56 & 2.224 & 1.299 & 12.4 & R & Schmidt\\
2653 Principia	    & 2004 Feb. 14 16:55-18:03 & 2.249 & 1.284 &  7.0 & R & 0.25m\\
(non Vesta family) & 2004 Feb. 15 14:44-19:52 & 2.249 & 1.280 &  6.5 & R & 0.25m\\
                   & 2004 Feb. 23 12:24-19:09 & 2.250 & 1.264 &  2.9 & R & K.3T\\
                   & 2004 Feb. 25 11:46-14:25 & 2.250 & 1.262 &  2.4 & R & K.3T\\
                   & 2004 Feb. 27 10:37-19:05 & 2.250 & 1.262 &  2.3 & R & K.3T\\
                   & 2004 Feb. 29 10:39-14:39 & 2.251 & 1.262 &  2.6 & R & 0.25m\\
2795 Lepage        & 2003 Sep. 27 16:12-18:22 & 2.315 & 1.474 & 17.0 & R & Schmidt\\
(non Vesta family) & 2003 Oct. 02 15:39-17:40 & 2.314 & 1.435 & 15.1 & R & Schmidt\\
                   & 2003 Oct. 04 16:45-17:35 & 2.313 & 1.420 & 14.3 & R & Schmidt\\
3307 Athabasca	    & 2004 Jan. 15 09:08-14:29 & 2.067 & 1.391 & 24.3 & R & Schmidt\\
(non Vesta family) & 2004 Jan. 16 10:43-10:58 & 2.068 & 1.391 & 24.3 & R & Schmidt\\
                   & 2004 Jan. 20 08:57-09:54 & 2.070 & 1.400 & 24.3 & R & Schmidt\\
                   & 2004 Jan. 21 11:40-12:13 & 2.071 & 1.443 & 24.5 & R & Schmidt\\
4147 Lennon        & 2003 Sep. 27 12:54-18:17 & 2.524 & 1.455 & 12.1 & R & Schmidt\\
(Vesta family)     & 2003 Sep. 28 18:44-19:30 & 2.525 & 1.609 & 11.6 & R & Schmidt\\
                   & 2003 Oct. 03 15:20-18:15 & 2.527 & 1.585 &  9.7 & R & Schmidt\\
                   & 2003 Oct. 04 16:40-17:30 & 2.527 & 1.580 &  9.3 & R & Schmidt\\
                   & 2003 Oct. 08 15:30-19:17 & 2.529 & 1.565 &  7.7 & R & Schmidt\\
4977 Rauthgundis   & 2004 Mar. 12 10:58-17:17 & 2.353 & 1.360 &  1.4 & R & Schmidt\\
(Vesta family)     & 2004 Mar. 13 10:30-17:45 & 2.352 & 1.358 &  0.9 & R & Schmidt\\
                   & 2004 Mar. 14 11:28-17:43 & 2.351 & 1.356 &  0.5 & R & Schmidt\\
                   & 2004 Mar. 15 11:53-16:34 & 2.350 & 1.355 &  0.5 & R & Schmidt\\
                   & 2004 Apr. 25 10:27-15:27 & 2.299 & 1.507 & 19.2 & R & Schmidt\\
                   & 2004 Apr. 28 10:41-15:12 & 2.295 & 1.531 & 20.2 & R & Schmidt\\
\noalign{\smallskip}
\tableline
\noalign{\smallskip}
%63 Ausonia (S)     & 2000 Dec. 29 09:25-17:54 & 2.684 & 1.803 & 11.4 & V & 0.25m\\
%238 Hypatia (C)    & 2003 Oct. 03 10:39-16:06 & 2.711 & 1.828 & 12.2 & V & K.3T\\
%                   & 2003 Oct. 04 11:08-15:51 & 2.710 & 1.835 & 12.5 & V & K.3T\\
%355 Gabriella (S)  & 2001 Nov. 24 12:25-16:24 & 2.281 & 1.358 & 11.4 & V & 0.25m\\
1455 Mitchella (A) & 2004 Apr. 29 11:02-18:19 & 2.000 & 1.162 & 11.0 & R & Schmidt\\
3192 A'Hearn (C)   & 2004 Mar. 12 09:45-10:48 & 1.984 & 1.039 & 12.4 & R & Schmidt\\
                   & 2004 Mar. 13 09:44-18:22 & 1.984 & 1.045 & 13.0 & R & Schmidt\\
                   & 2004 Mar. 14 09:46-17:02 & 1.985 & 1.049 & 13.4 & R & Schmidt\\
                   & 2004 Mar. 15 09:47-17:56 & 1.985 & 1.055 & 14.0 & R & Schmidt\\
6664 Tennyo (?)    & 2004 Jan. 15 17:05-20:59 & 2.238 & 1.426 & 17.9 & R & Schmidt\\
\noalign{\smallskip}
\tableline
\end{tabular}
}
\end{center}
\end{table}

   Reduction of obtained images reduction including dark subtraction and flat field 
correction was carried out by the Image Reduction and Analysis Facility (IRAF) software.
   Measurements of the asteroids and the comparison stellar stars were done thorough a 
circular aperture with a diameter of 3 times of seeing size using the APPOT task of IRAF.
   This aperture size was chosen in such a way as to collect about more than 99 $\%$ of 
the scattered light from the objects.  
   Fluxes of the asteroids were measured relative to the on-chip 5-15 comparison stars 
in each asteroid frame.  
   Attention was paid to check all comparison stars.
   If a comparison star is a variable star, the star was no adopted as comparison stars.
   Brightness of typical comparison stars were brighter than that of the asteroid.

   To obtain accurate lightcurve data, different of the light-travel time was taken 
into account.
   When one part of images observed on different days overlapped, lightcurve data 
were reduced to unit the heliocentric and geocentric distance and phase function.  
   As the G parameter of V-type asteroids and others for phase function correction 0.32 
(G parameter value of Vesta) and 0.15 (G parameter value of typical asteroids) were 
employed.
   It is empirically known that the amplitude of asteroids is changed by phase angle 
(Zappala et al. 1990). 
   Zappala et al. (1990) suggested that the coefficient m varies with asteroidal 
spectral types, but Ohba et al. (2003) showed that the coefficient m is related to not 
asteroidal spectral type but surface roughness.
   For correction of lightcurve amplitude at phase angle 0$\deg$, we adopted the 
coefficient m = 0.02 value was derived by median values of surface roughness of 
asteroids (Bowell et al. 1989).  
   To determine the asteroidal rotational periods, two methods: the Fourier analysis 
method (FFT) and the phase dispersion minimization method (PDM)(Stellingwerf 1978) were 
used.
   Rotation periods were determined under two methods.  
   In case of relative photometry (all data of 2653 Principia, between March and 
April data for 2511 Patterson and 4977 Rauthgundis), we did several iteration 
after the value of the lightcurve of each data were shifted.

\begin{figure}[!ht]
\plottwo{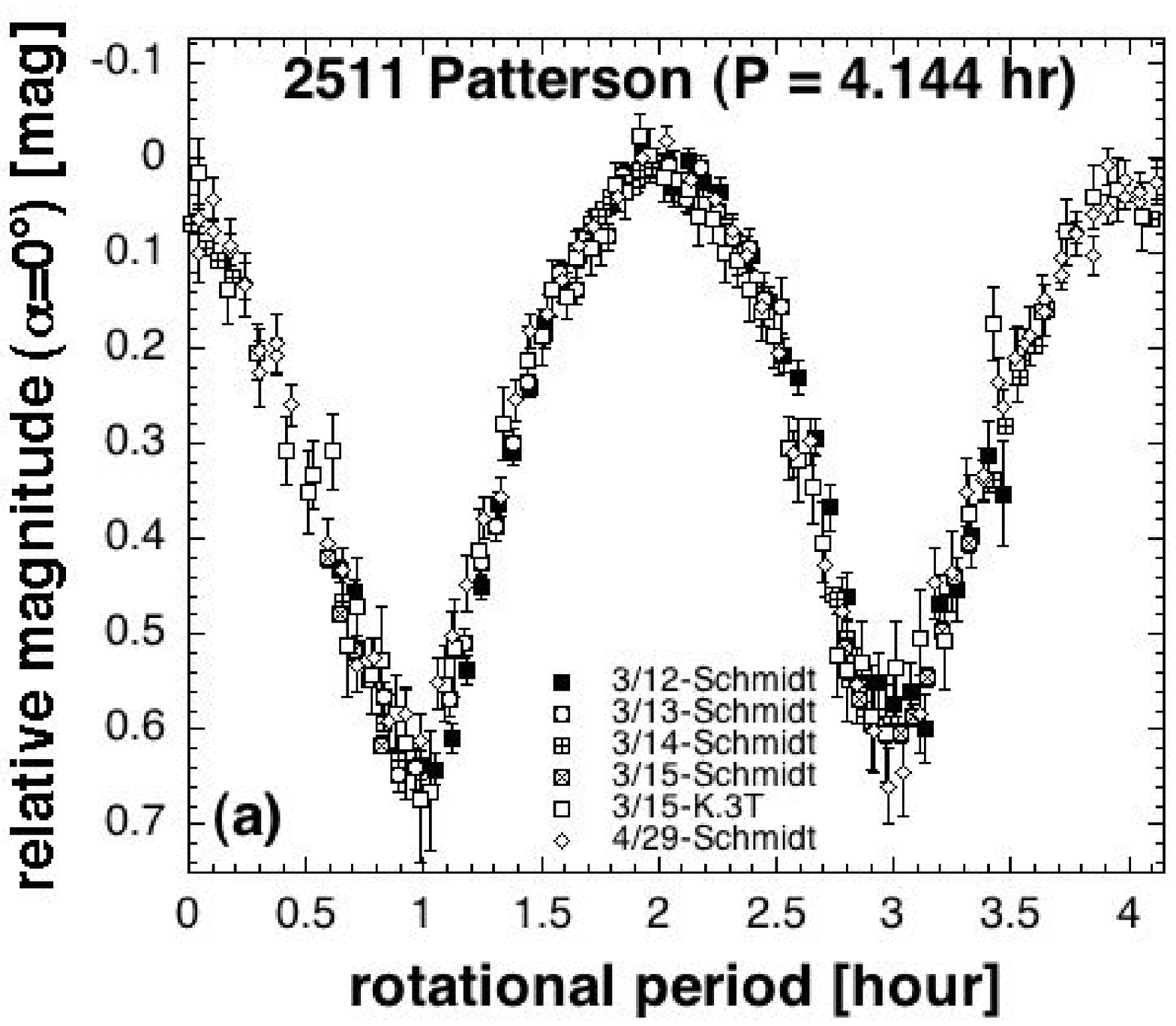}{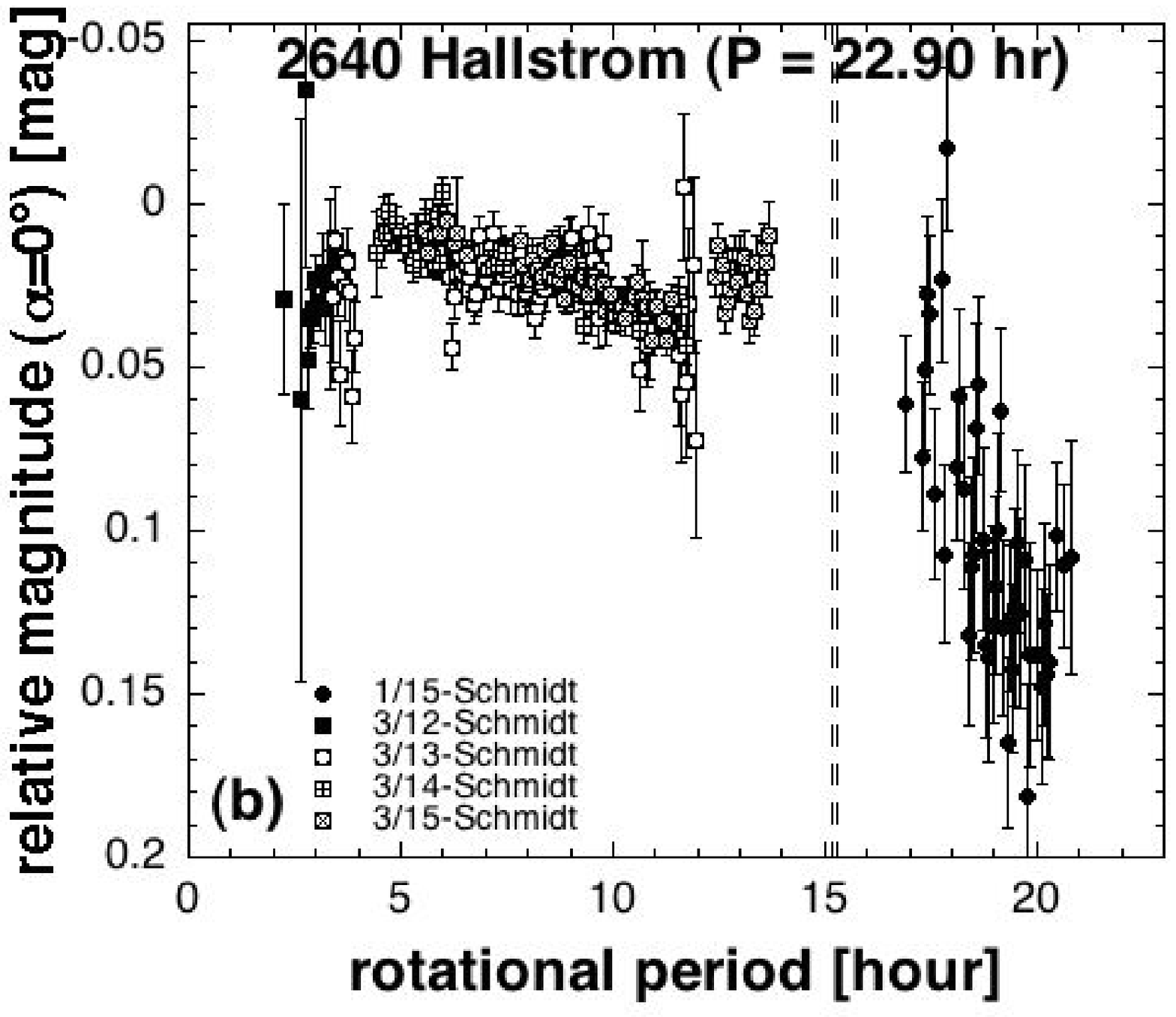}
\plottwo{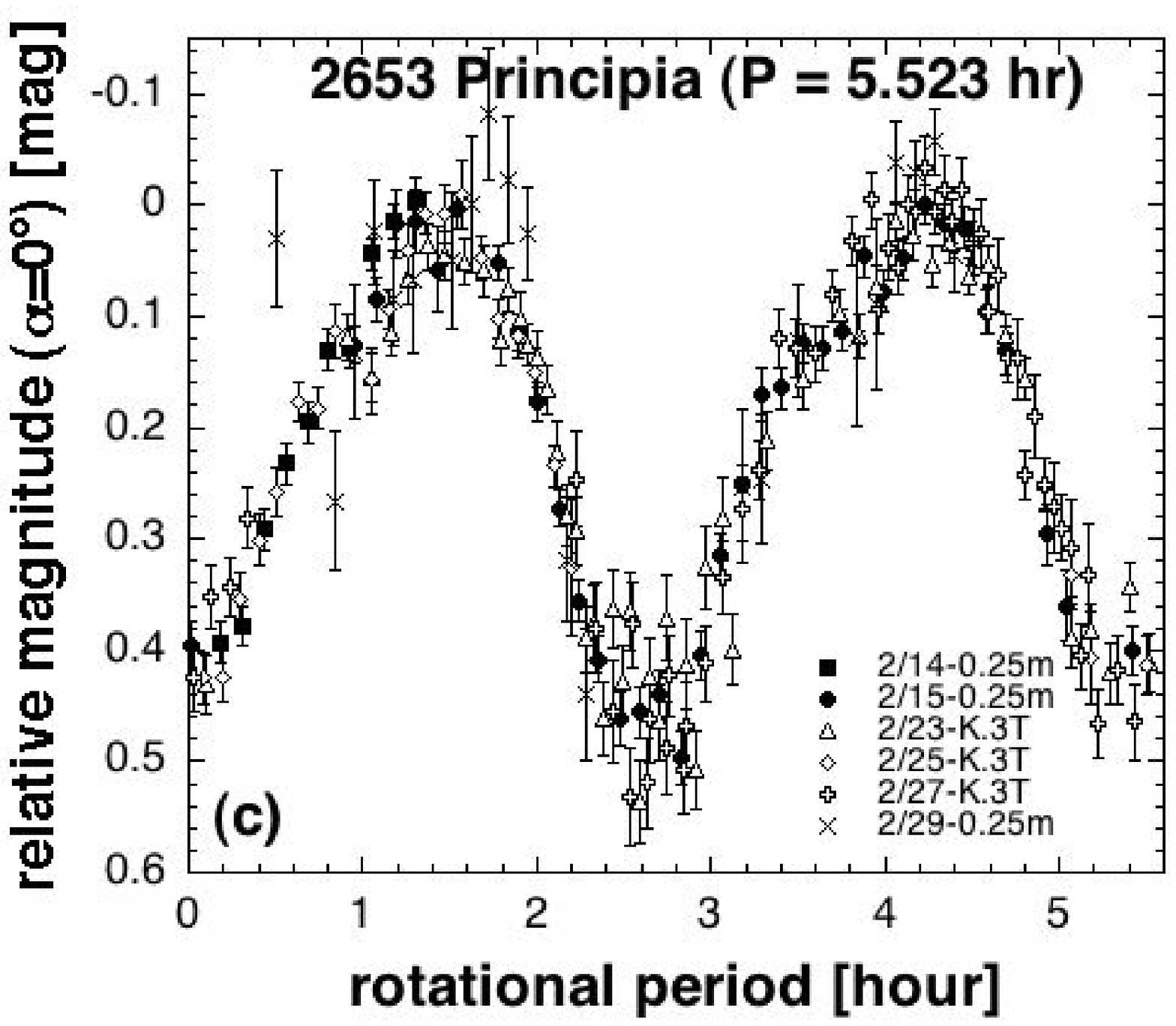}{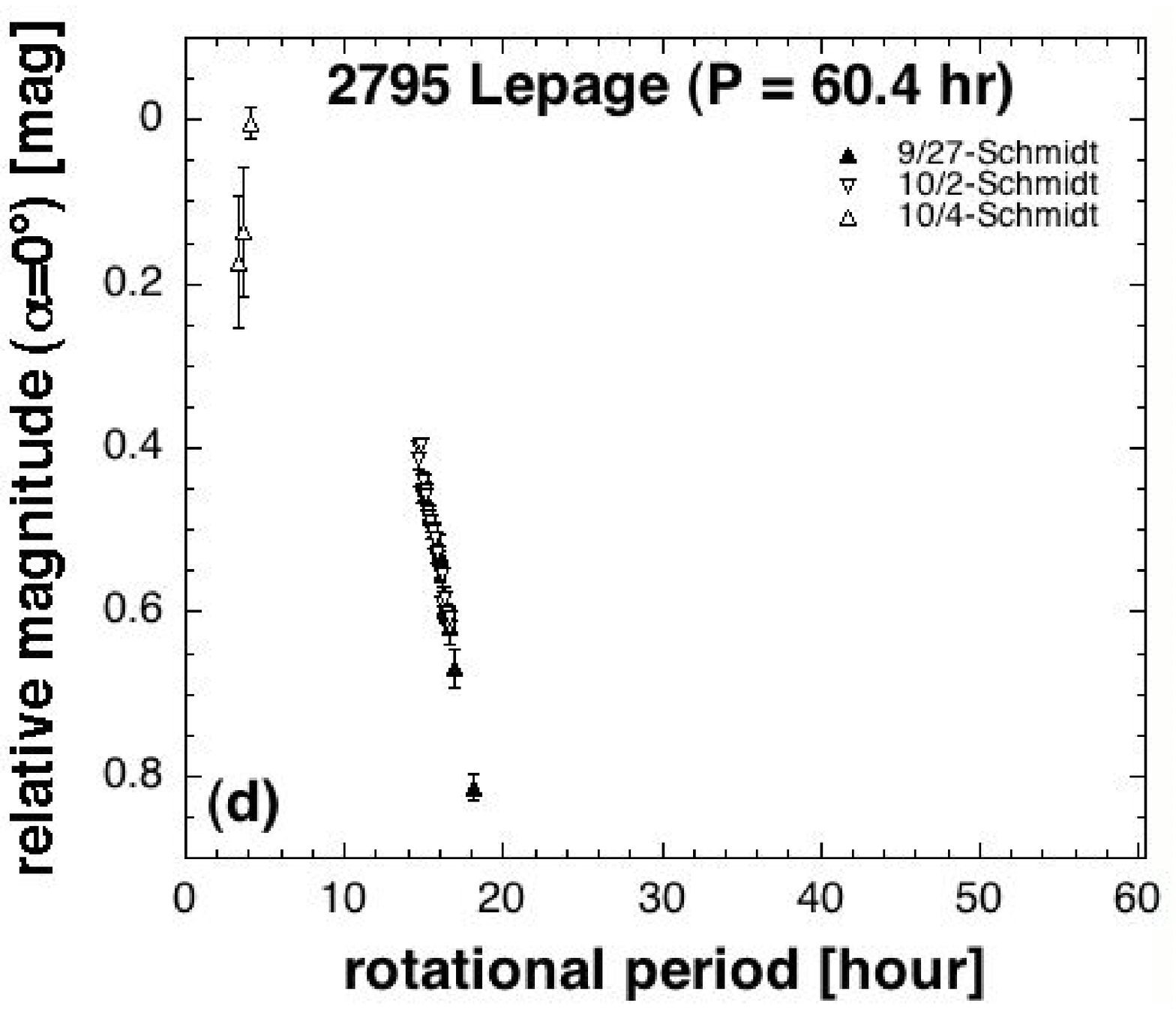}
%\plottwo{3307.eps}{4147.eps}
%\plottwo{4977.eps}{4977-kara.eps}
\caption{CCD lightcurves of V-type asteroid 2511 Patterson, 2640 Hallstorm, 2653 
Principia, and 2795 Lapage.}
\end{figure}

\section{Observational results}   %%% Top level section head (remove "%" symbol)
\subsection*{V-type asteroids}   %%% Unnumbered second level section head (remove "%" symbol)
\subsubsection*{2511 Patterson (Fig. 1a)}   %%% Lowest level section head (remove "%" symbol)
   The asteroid 2511 Patterson is a V-type asteroid (Bus \& Binzel 2002) and a member 
of the Vesta family (Zappala et al. 1995).  
   Assuming its albedo to be the same as Vesta's albedo (pv = 0.36), we estimated the 
diameter of the asteroid to be $\sim$7 km.
   The period of this asteroid was not known before this study. 
   The asteroid was made the observation using two telescopes at the Kiso observatory 
for five nights in the R band (Hasegawa et al. 2004).
   The asteroid has a symmetric lightcurve with an amplitude of 0.7 mag. 
   A period of 4.144 $\pm$ 0.001 hours was uniquely determined by FFT and PDM methods.  

\subsubsection*{2640 Hallstrom (Fig. 1b)}   %%% Lowest level section head (remove "%" symbol)
   The asteroid 2640 Hallstrom is a Vestoid (Bus \& Binzel 2002) but is not a member 
of the Vesta family (Zappala et al. 1995).
   However, the asteroid is positioned between asteroid Vesta and the 3:1 mean resonance 
of Jupiter.  
   If its albedo is the same as that of Vesta (pv = 0.36), the diameter of the asteroid 
is $\sim$5.5 km.
   The period of this asteroid was not known before this study. 
   The lightvurve data were taken for five nights in the R band.
   The lightcurve of the asteroid has an asymmetric shape. 
   We found a rotation period to be equal to 22.90 $\pm$ 0.05 hours using FFT and PDM 
methods.  

\subsubsection*{2653 Principia (Fig. 1c)}   %%% Lowest level section head (remove "%" symbol)
   The 2653 Principia which has a Vesta-like spectrum (Bus \& Binzel 2002) does not 
belong to the Vesta family (Zappala et al. 1995).
   Based on an assumed albedo of 0.36, the diameter of the asteroid was estimated 
to be $\sim$8.5 km .
   The asteroid was observed in six nights with the 0.25-m telescope and the K.3T 
(Hasegawa et al. 2004).
   Our obtained lightcurve of the asteroid is slightly asymmetric with maximum and 
minimum which differ from each other by shape.
   A solution of 5.522 $\pm$ 0.005 hours was determined from FFT and PDM methods.
   Most at the same time, the asteroid has been also observed using 0.37-m F/14 
Cassegrain reflector at the Iowa Robotic Observatory (MPC code 857) (Willis 2004) and 
using 0.81-m F/7 Ritchey-Chretien telescope at the Tenagra observatory (Windschmitl 
\& Vonk 2004).  
   They reported the rotational period of 5.523 hours and 6.243 hours, which is in a 
good agreement with our result.

\subsubsection*{2795 Lepage (Fig. 1d)}   %%% Lowest level section head (remove "%" symbol)
   The asteroid 2795 Lepage is a V-type asteroid (Bus \& Binzel 2002) but is not 
a member of the Vesta family (Zappala et al. 1995).  
   If its albedo is the same as that of Vesta (pv = 0.36), its diameter is $\sim$5 km.
   The period of this asteroid was not known before this study. 
   The lightcurve observations were carried out for three nights in R band.
   Only small parts of lightcurve was taken, we found its rotatinal peroid P 60.4 
$\pm$ 0.5 hours using FFT and PDM methods.  

\begin{figure}[!ht]
\plottwo{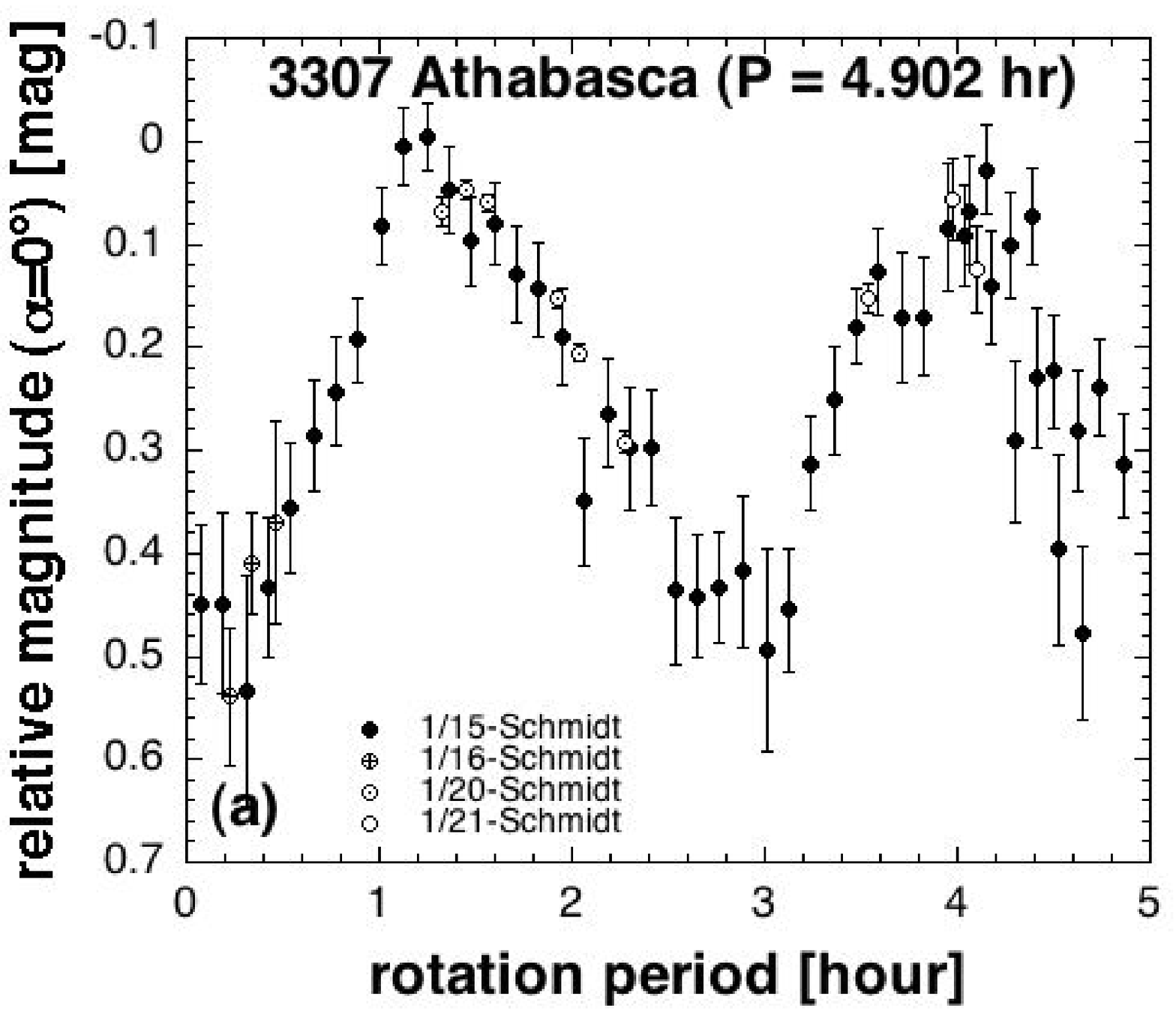}{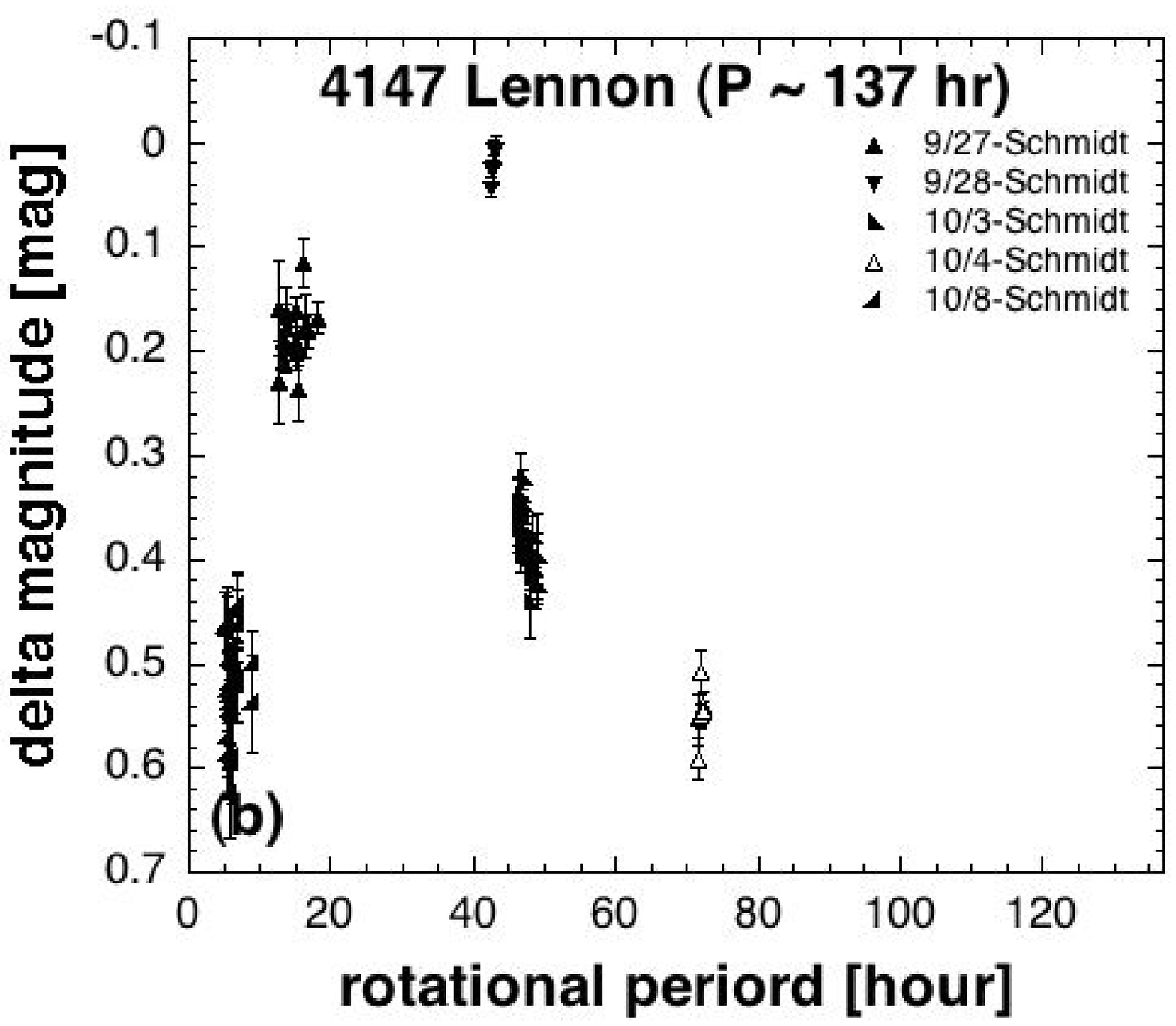}
\plottwo{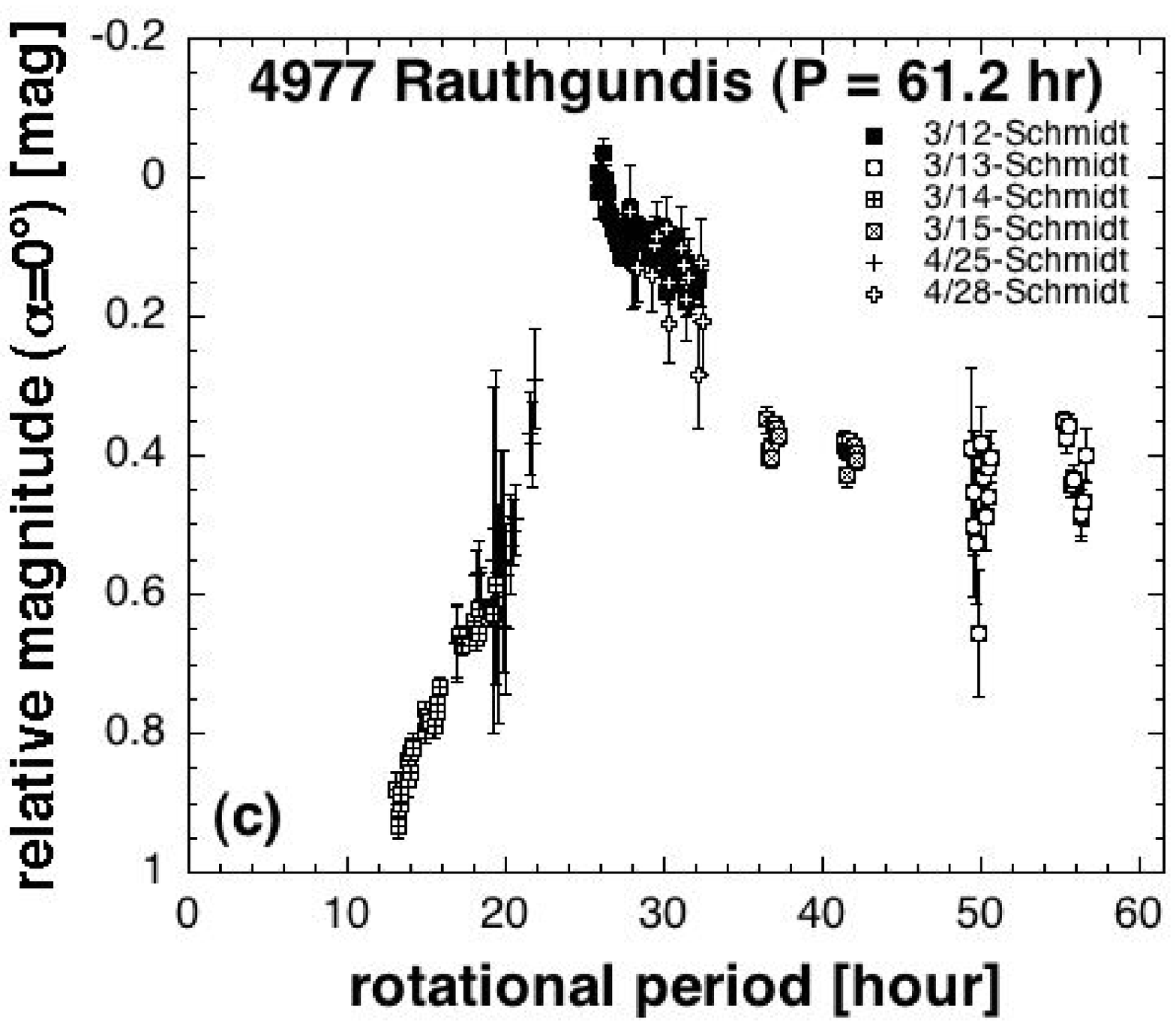}{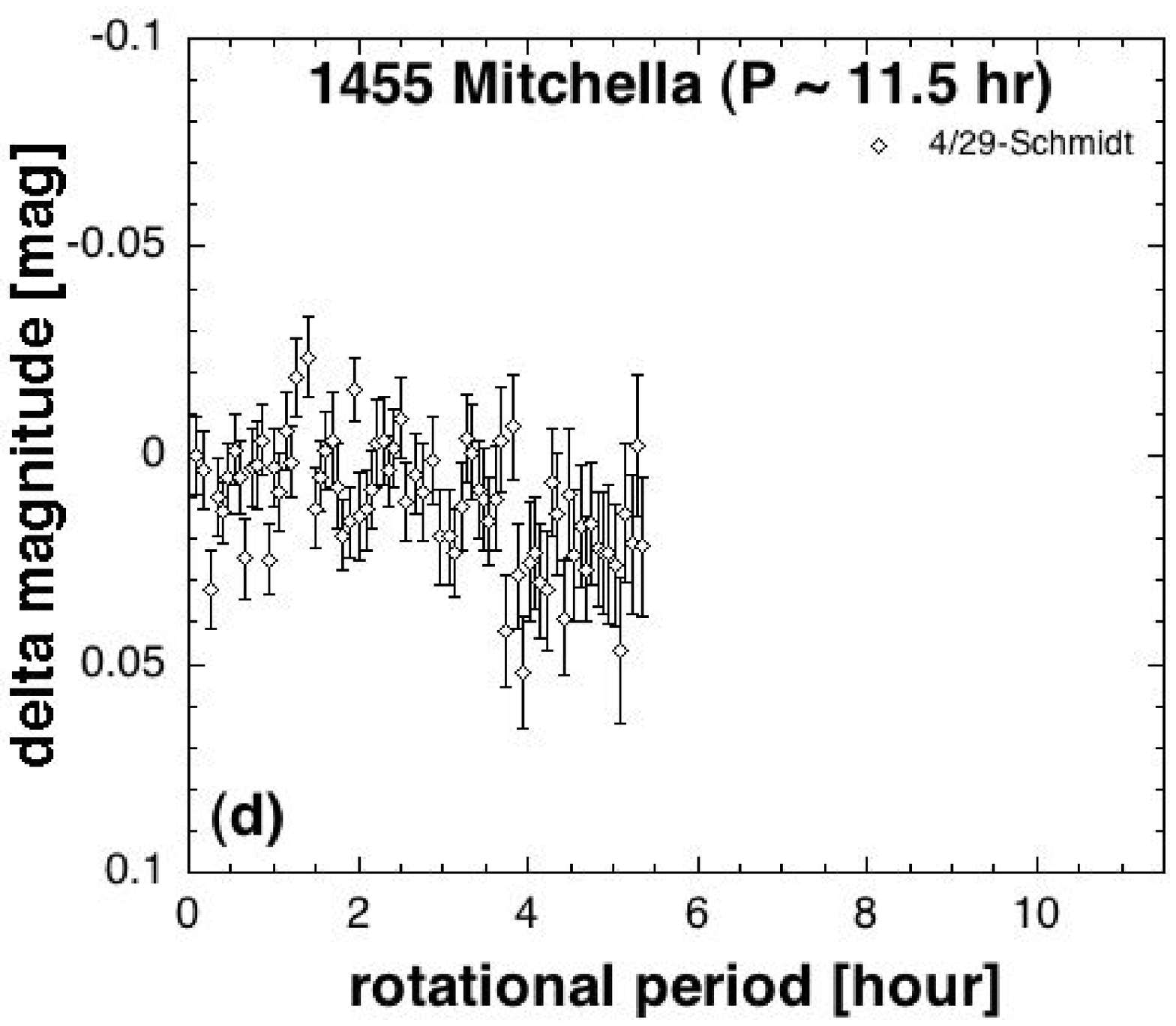}
\plottwo{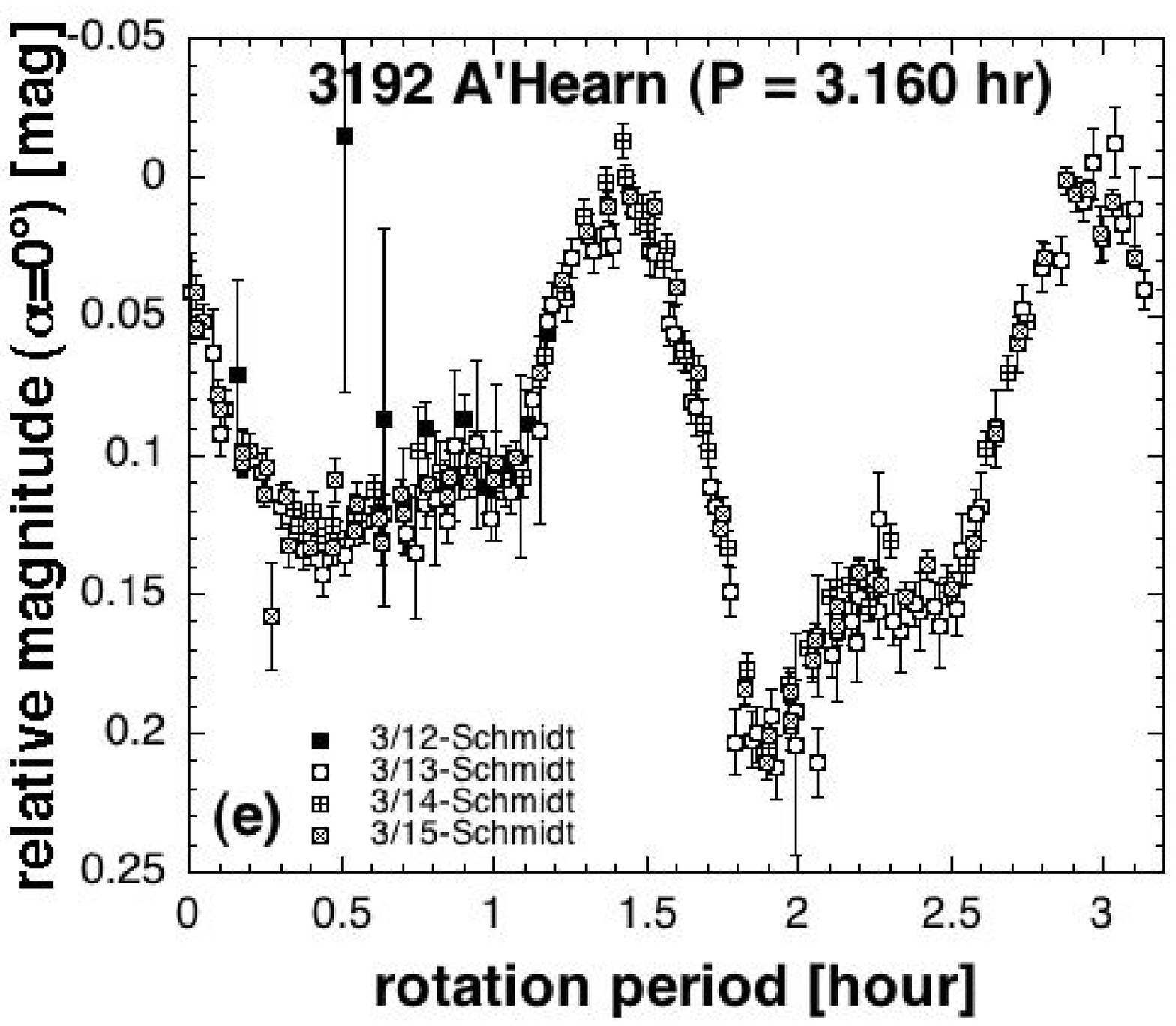}{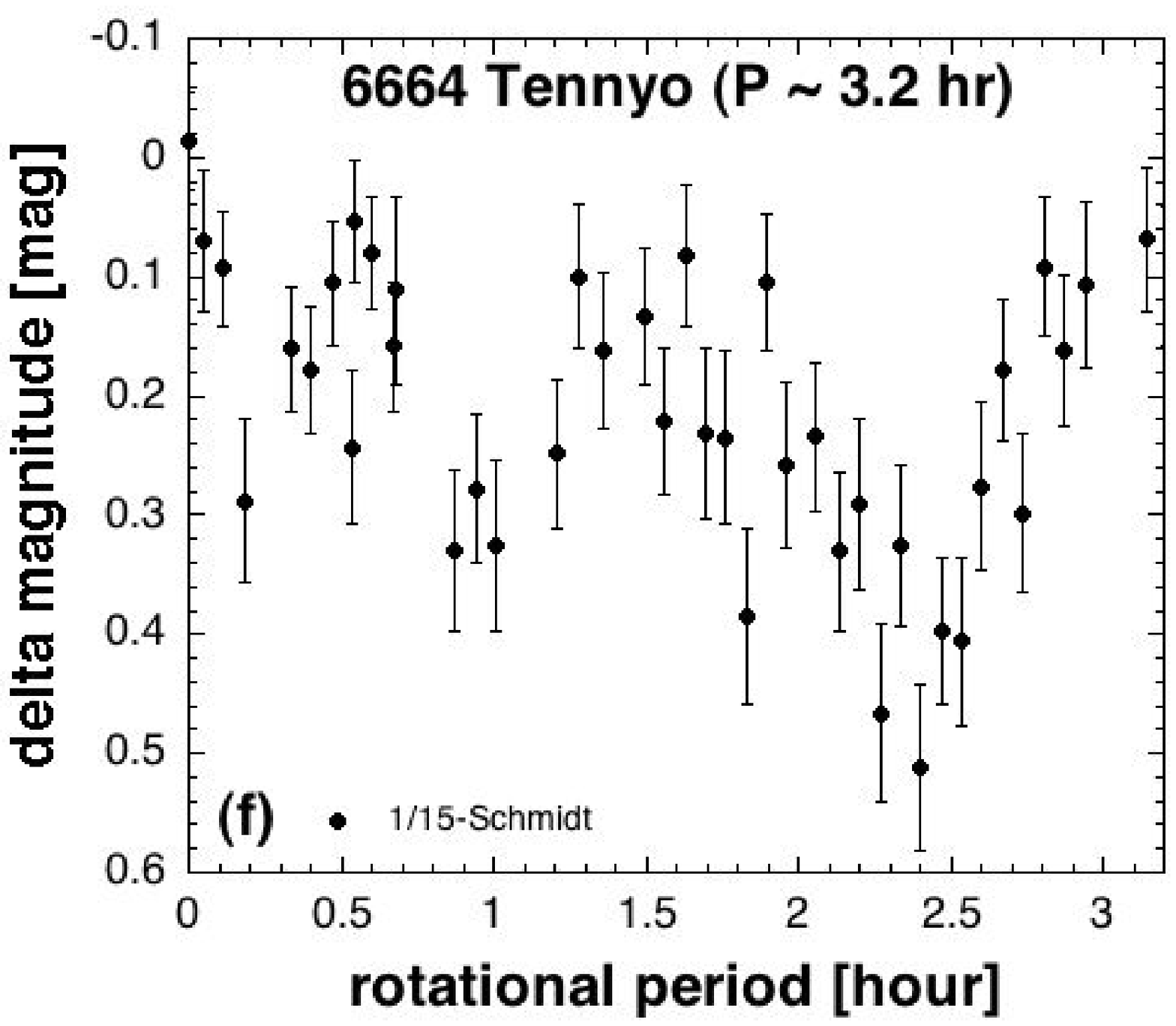}
\caption{CCD lightcurves of V-type asteroid 3307 Athabasca, 4147 Lennon, and 4977 
Rauthgundis, and non V-type asteroid 1455 Mitchella (A-type), 3192 A'Hearn (C-type), 
and 6664 Tennyo (spectral type unknown)
}
\end{figure}

\subsubsection*{3307 Athabasca (Fig. 2a)}   %%% Lowest level section head (remove "%" symbol)
   The asteroid 3307 Athabasca is a Vestoid (Bus \& Binzel 2002) and a member of the 
Vesta family (Zappala et al. 1995).  
   Assuming its albedo to be the same as Vesta's albedo (pv = 0.36), the size of the 
asteroid is $\sim$3.5 km in diameter.
   The period of this asteroid was not known before this study. 
   The asteroid was observed for four nights in the R band.
   The lightcurve shape of the asteroid is almost symmetric with a maximum amplitude 
 of 0.5 mag. 
   We have determined the rotation period of this asteroid to 4.902 $\pm$ 0.010 hours 
by FFT and PDM methods.  

\subsubsection*{4147 Lenonn (Fig. 2b)}   %%% Lowest level section head (remove "%" symbol)
   The 4147 Lenonn which has a Vesta-like spectrum (Xu et al. 1993) belongs to the Vesta 
family (Zappala et al. 1995). 
   The diameter of the asteroid was estimated to be $\sim$5.5 km with the assumption
that its albedo is the same as Vesta'a albedo.
   The period of this asteroid was not known before this study. 
   The asteroid was monitored for five nights in the R band.
   Although we have obtained only small parts of lightcurve data, a period of $\sim$137 
hours is suggested by only FFT methods.  

\subsubsection*{4977 Rauthgundis (Fig. 2c)}   %%% Lowest level section head (remove "%" symbol)
   The asteroid 4977 Rauthgundis is a V-type asteroid (Bus \& Binzel 2002) and 
a member of the Vesta family (Zappala et al. 1995).  
   If its albedo is the same as that of Vesta (pv = 0.36), the asteroid is $\sim$4 km
in diameter.
   The period of this asteroid was not known before this study. 
   The lightcurve observation of the asteroid was performed for six nights in the R band.
   The lightcurve of the asteroid has an asymmetric shape with a maximum amplitude more 
than 0.9 mag. 
   A solution of 61.2 $\pm$ 0.3 hours was derived from FFT and PDM methods. 

\subsection*{non V-type asteroids}   %%% Unnumbered second level section head (remove "%" symbol)

\subsubsection*{1455 Mitchella (Fig. 2d)}   %%% Lowest level section head (remove "%" symbol)
   The asteroid 1455 Mitchella is a A-type asteroid  (Lazzaro et al. 2004). 
   The period of this asteroid was not known before this study. 
   The asteroid was monitored in the same field as 2511 Patterson for five nights 
in the R band.
   Since we have obtained only a part of lightcurve data, a period of $\sim$11.5 
hours is suggested.  

\subsubsection*{3192 A'Hearn (Fig. 2e)}   %%% Lowest level section head (remove "%" symbol)
   The asteroid 3192 A'Hearn is a C-type asteroid  (Bus \& Binzel 2002). 
   The period of this asteroid was not known before this study. 
   The asteroid was observed in the same frame as 2640 Hallstrom for consecutive 
four nights in the R band.
   The lightcurve of the asteroid has an asymmetric shape with a maximum amplitude more 
than 0.2 mag. 
   We found rotation period to be equal to 3.160 $\pm$ 0.010 hours using FFT and 
PDM methods. 

\subsubsection*{6664 Tennyo (Fig. 2f)}   %%% Lowest level section head (remove "%" symbol)
   Spectral type of the asteroid 6664 Tennyo is not known. 
   The period of this asteroid was not known before this study. 
   The asteroid was observed in the same frame as 2640 Hallstrom for one nights 
in the R band.
   The asteroid has an asymmetric lightcurve with a maximum amplitude of $\sim$0.4 mag. 
   The period was suggested to be $\sim$3.2 hours.

\section{Discussion}   %%% Top level section head (remove "%" symbol)

  We have shown rotational rates for V-type asteroids by obtaining new data and using 
past data on rotational periods. 
  Past data were obtained from Harris' lightcurve data catalogue on below web site: 
http://cfa-www.harvard.edu/iau/ lists/LightcurveDat.html.
  The distribution of rotational rates for V-type near-Earth asteroids (NEAs) has a 
single peak (Fig. 3a), but we found triple peaks in the distribution of spin rates for 
V-type main-belt asterois(MBAs) (Fig. 3b). 
  The distribution of spin rate for observed V-type MBAs of which absolute magnitudes 
(Hmag) distribute by the range from 11.5 to 15.0 mag (Fig. 3c) is not similar to that 
for all asteroids with 11.5 $\le$ Hmag $\le$ 15.0 mag.

\begin{figure}[!ht]
\plottwo{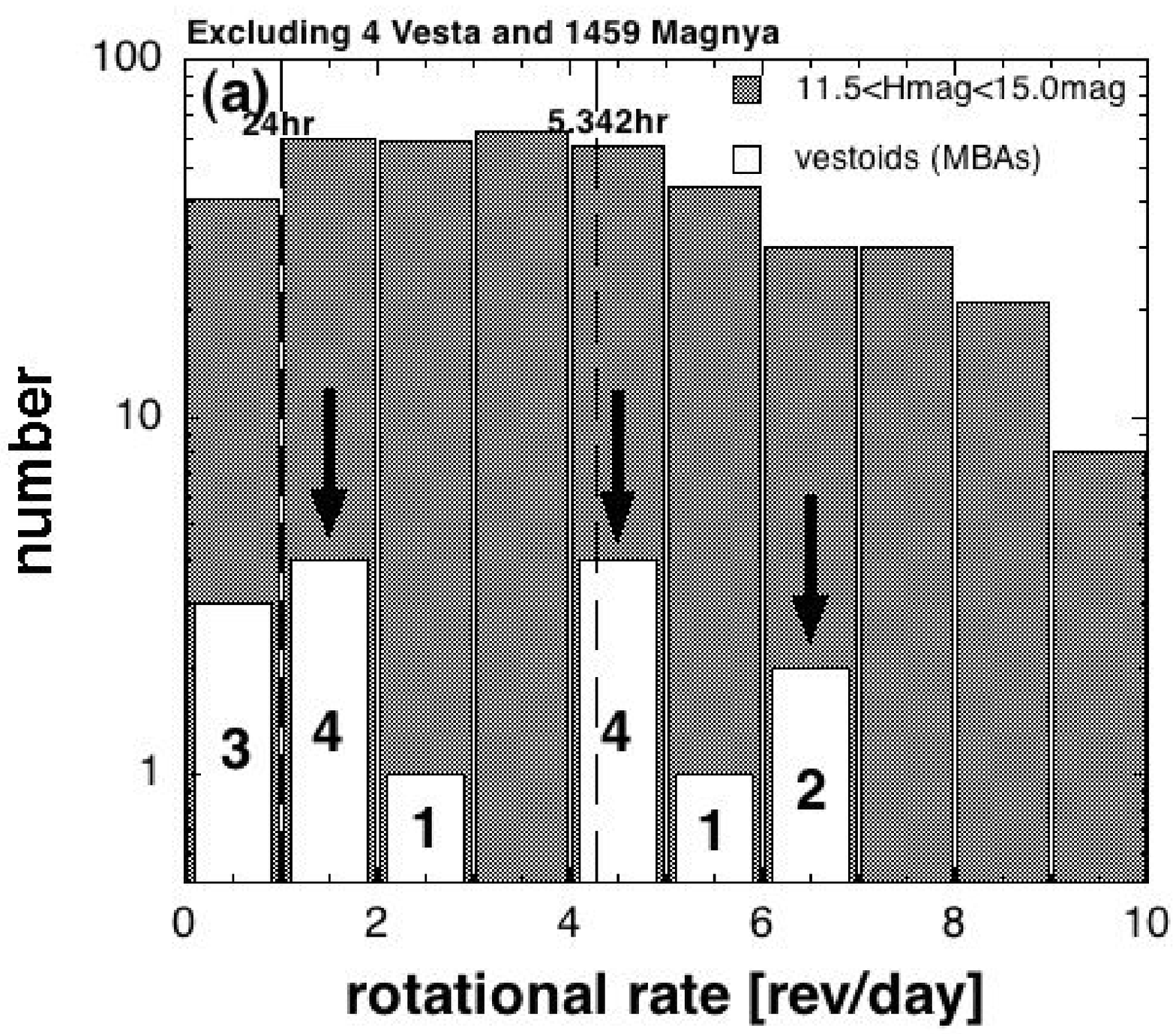}{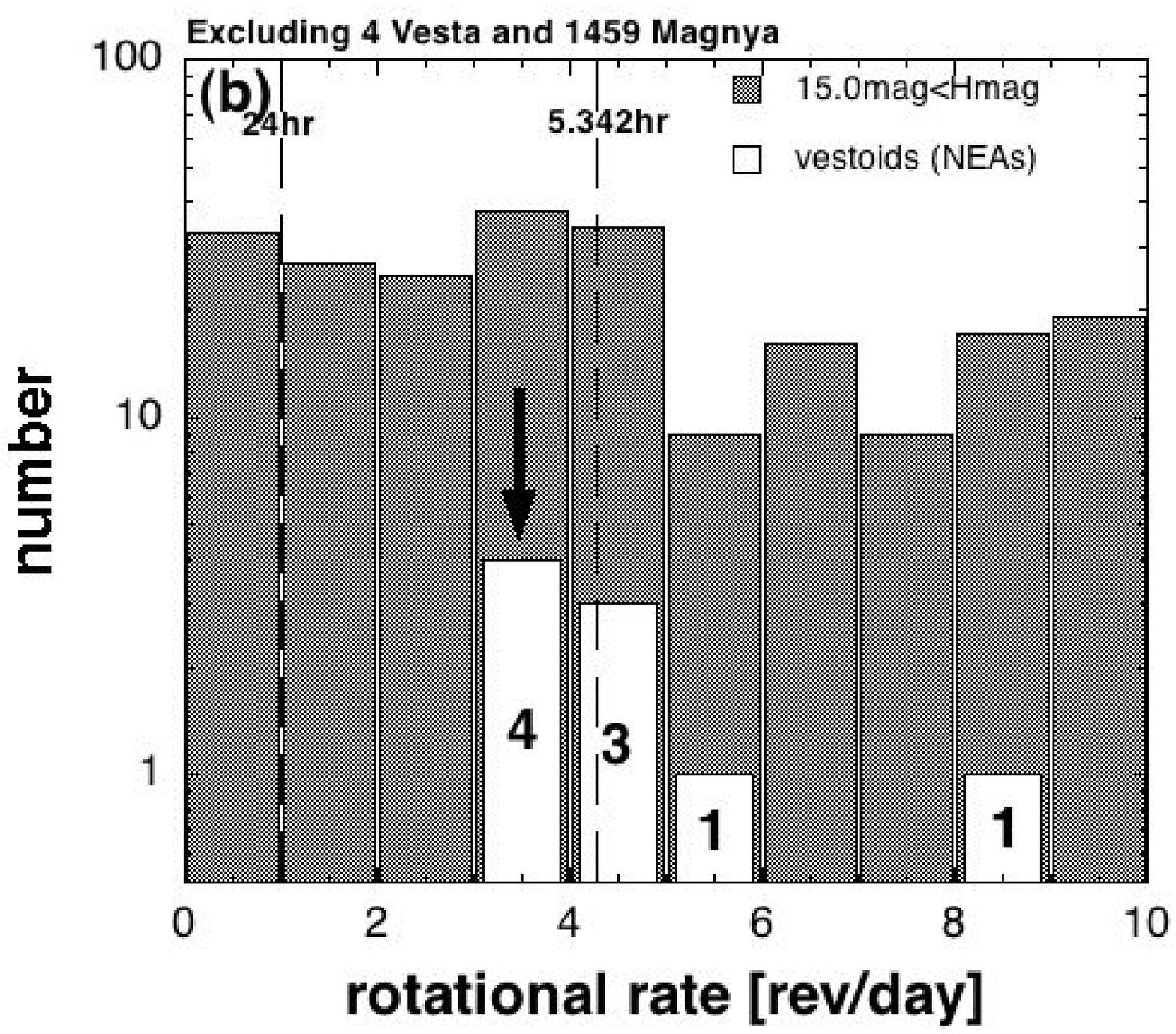}
\plottwo{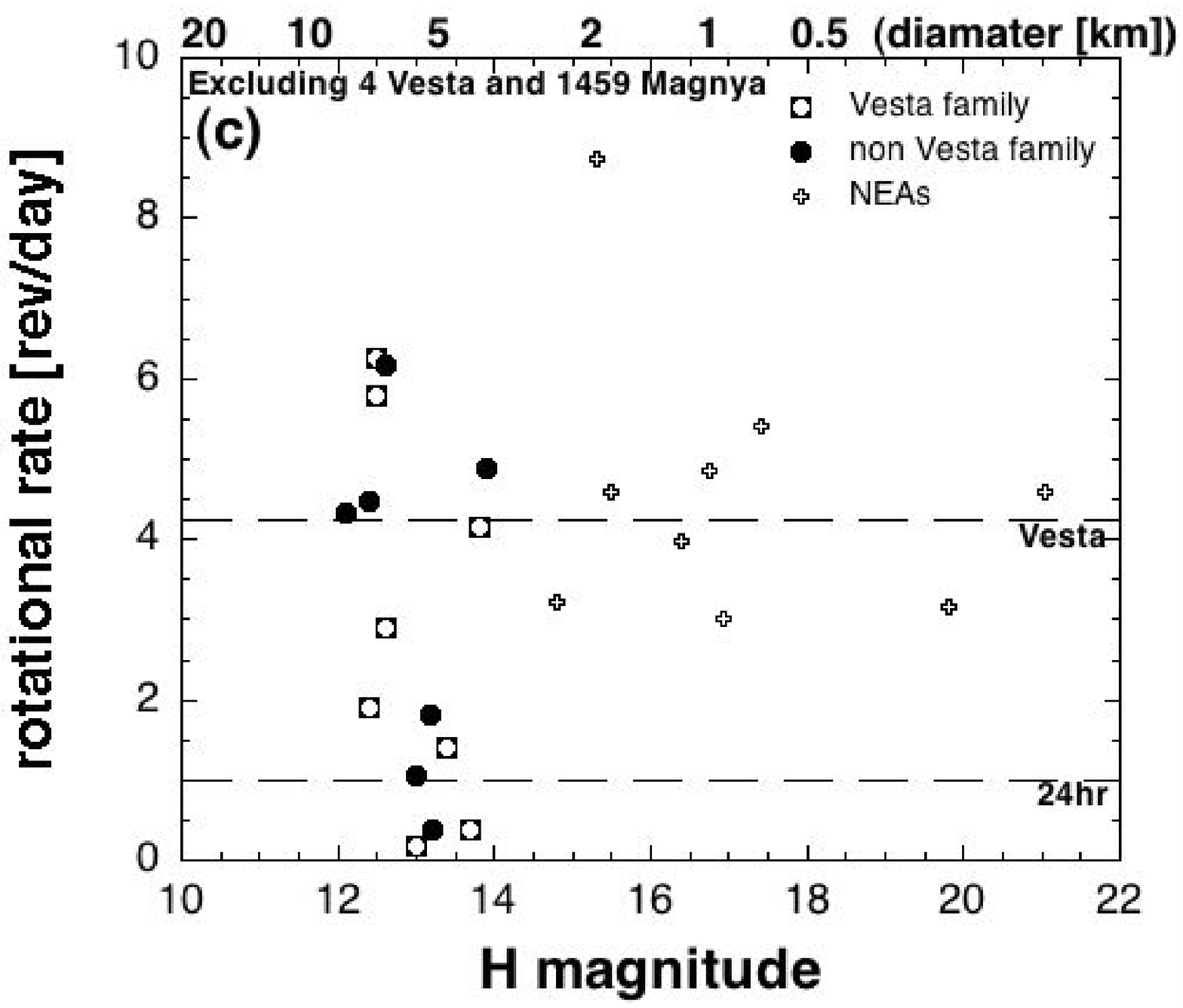}{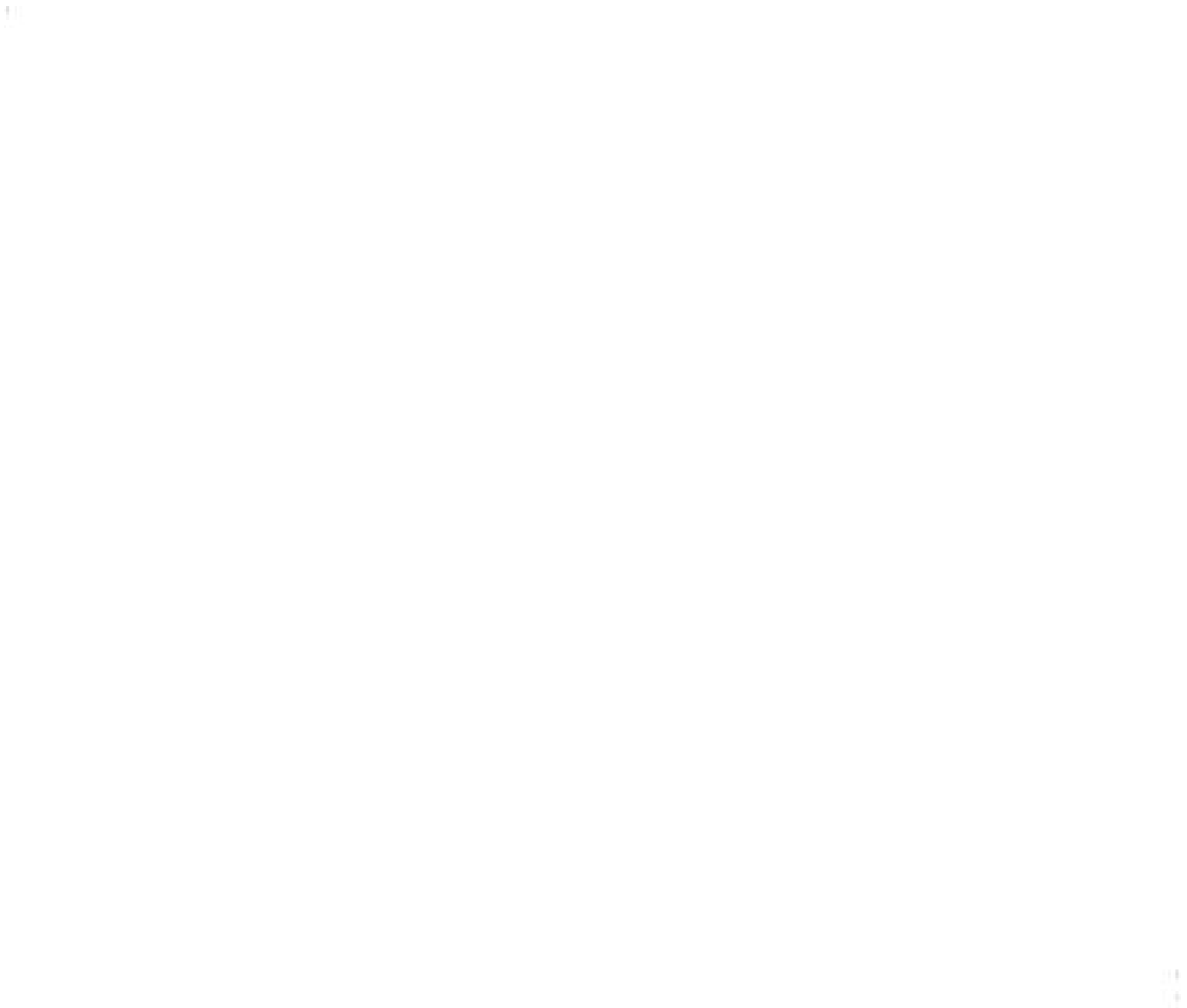}
\caption{{\itshape a:\/} Histgram of rotational rates for V-type MBAs and all 
spin rate known asteroids with the range from 11.5 to 15.0 magnitude.
{\itshape b:\/} Histgram of rotational rates for V-type NEAs and all 
spin rate known asteroids below 15.0 magnitude.
{\itshape c:\/} Asteroidal rotational rate via absolute magnitudes for V-type 
asteroids.}
\end{figure}

  Welton et al. (1997) showed that cosmic-ray-exposure ages (CREAs) of HED meteorites 
is the range from 6 to 50 Ma.
  This time scale is the same level of collisional time of young families such as Karin, 
Veritas, and Iannini families (Nesvorny et al. 2003).
  These families have dust bands.
  This is not contradictory to the fact that the dynamical lifetime of micron to sub-mm 
sized particles is believed to be $\sim$0.1-1 Myr.
  However, the Vesta family does not have any associated dust bands.
  This fact implies that the Vesta family is ancient.
  From a point of view of dynamics, it is consistent that HED meteorites are ejecta 
which not come from Vesta directly but went out of surface of V-type asteroids such as
V-type NEAs.

  Marzari et al (1996) also indicated that a large impact forming Vesta family occured 
$\sim$1 Gyr by dynamical simulation. 
  On the other hand, Yamaguchi et al, (2001) suggested that the partial melting event 
that reset the ages $\sim$4.50 Ga ago was caused by an impact into the hot crust of 
Vesta from mineralogic, radiometric, and ion microprobe reseaches of the eucrite.
  Wakefield (2004) also suggested the scenario that some of V-type asteroids which are 
parent bodies of HED meteorites were ejected $\sim$ 3.5 Gyr ago from the study combined 
CREAs and Ar-Ar ages of eucrites.
  
   Vorkrouhlicky et al. (2003) found that prograde rotators have 
2-3 revolution per day and retrograde rotators have less than 2 and more than 5 
revolution per day for 20-40 km sized Koronis family asteroids which are considered 
to have been formed sevearl Gyr ago.
  They pointed out that spin rates of the asteroids which was formed by collision of 
$\sim$Gyr ago were affected by the YORP effect.
  It is possible that three peaks in distribution of V-type MBAs is caused by the 
YORP effect.
  This implies the formation age of the Vesta family is not young but old. 

\acknowledgements We thank the Kiso observatory staffs for their support of lightcurve 
observations.
  We are greatly indebted to Dr. T. Hiroi and Dr. S. Sasaki for their useful comments.

%%% Text of acknowledgements runs on after this command.

%%% THE BIBLIOGRAPHY
%%%
%%% CONSULT SECTION 3 OF "INSTRUCTIONS FOR AUTHORS" FOR HOW TO USE NATBIB.
%%% AUTHORS ARE ENCOURAGED TO USE EITHER THE "THEBIBLIOGRAPY" ENVIRONMENT
%%% BY UNCOMMENTING (DELETING THE "%" SYMBOL) THE COMMANDS BELOW, OR BY
%%% USING THE BIBTEX ENVIRONMENT. TO FIND OUT WHICH IS APPLICABLE TO YOUR
%%% CONTRIBUTION, CONSULT THE VOLUME EDITORS FOR YOUR PROCEEDINGS.
%%%

\end{document}